\documentclass{article}
\usepackage[utf8]{inputenc}
\usepackage{authblk}
\usepackage{setspace}
\usepackage[margin=1.25in]{geometry}
\usepackage{graphicx}
\graphicspath{ {./figures/} }
\usepackage{subcaption}
\usepackage{amsmath}
\usepackage{lineno}
\usepackage{gensymb}
\usepackage{amssymb}
\usepackage{booktabs}
\usepackage{multirow}
\usepackage{hyperref}
\hypersetup{
    hidelinks,          
    pdfstartview=FitH   
}

\usepackage[style=nejm, 
citestyle=numeric-comp,
sorting=none]{biblatex}
\title{Simulation and Deep Learning Inversion for Energetic Neutral Atom Imaging at Heliospheric Boundary}

\author[1]{Weining Wang}
\author[1]{Rong Lin}
\author[*1,2]{Jiansen He}
\author[*1,2]{Hong Zou}
\author[1]{Yongfu Wang}
\author[1]{Jiali Chen}
\author[1,3]{Yuguang Ye}
\author[1]{Weihong Shi}
\author[4]{Wei Zhang}
\author[4]{Xi Lu}
\author[5]{Die Duan}
\author[1]{Tianhang Chen}
\author[1,2]{Linghua Wang}
\author[1,3]{Qiugang Zong}

\affil[1]{School of Earth and Space Sciences, Peking University, Beijing, China.}
\affil[2]{State Key Laboratory of Solar Activity and Space Weather, National Space Science Center, Chinese Academy of Sciences, Beijing, China.}

\affil[3]{State Key Laboratory of Lunar and Planetary Sciences, Macau University of Science and Technology, Macau, China.}
\affil[4]{Shanghai Institute of Satellite Engineering,  Shanghai, China.}
\affil[5]{College of Advanced Interdisciplinary Studies,
National University of Defense Technology,
Changsha, China}

\affil[*]{Address correspondence to: jshept@pku.edu.cn and hongzou@pku.edu.cn}

\date{}

\begin{document}

\maketitle

\begin{abstract}
The distribution of energetic neutral atom (ENA) differential flux density at the heliospheric boundary is the key to revealing the interaction mechanism between the solar wind and the interstellar medium. The Helios ENA Rapid Mapper (HERMA) is a pre-research project for a new ENA detection concept and technical principle, which is designed to provide technical support for refined detection with its advantages of wide energy range coverage and high spatial resolution. To verify the high spatiotemporal resolution rapid imaging capability of the HERMA, this study restored the particle response characteristics of the HERMA (spatial modulation grid + 256-pixel linear array silicon detector) based on Geant4 simulation, and we compared the performance of the traditional inversion algorithm with the convolutional neural network (CNN) inversion algorithm using INCA data as the benchmark. The results show that under different sampling times, the source distribution inverted by the CNN algorithm has a higher degree of agreement with the ideally expected distribution, and the CNN algorithm exhibits better stability and accuracy than the traditional algorithm under the conditions of low signal-to-noise ratio and short staring time. This study provides core technical support for the engineering verification and on-orbit application of the HERMA instrument, fills the technical gap in wide-energy-range and high-resolution ENA detection, and helps to deepen the research on the interaction mechanism between the solar wind and the interstellar medium at the heliospheric boundary.
\end{abstract}


\section{Introduction}
\par
\textbf{The heliosphere is the region in space governed} by the solar wind and the magnetic field it carries. Solar wind (SW) plasma propagates antisunward with a supersonic speed and eventually decelerates under the interaction with the local interstellar medium (LISM) plasma, eventually becoming subsonic at the termination shock (TS). Beyond the TS, the subsonic and heated solar wind plasma continues to flow in the inner heliosheath (IHS) and interacts with the LISM plasma. This interaction forms the heliopause (HP), a tangential discontinuity that demarcates the heliosphere from the interstellar medium\cite{izmodenov2009local,mccomas2021interstellar,zank2013heliospheric,burlaga2019magnetic}. However, uncharged interstellar neutral atoms (mostly hydrogen atoms) are not affected by electromagnetic fields and pass through these structures unimpeded into the heliosphere. A portion of these neutral components undergoes charge exchange with the solar wind plasma\textbf{or photo-ionization due to solar UV} to become pickup ions (PUIs), which are captured by the solar wind and thus move outward. Some of the PUIs are heated and accelerated by interplanetary shocks, TS or other processes to become suprathermal PUIs with higher energy than solar wind thermal protons\cite{mccomas2017interstellar,mccomas2021interstellar,shrestha2024suprathermal,wang2025spectral,zirnstein2018situ}. This part of high-energy PUIs can exchange charge with neutral atoms again to become energetic neutral atoms (ENAs).
ENAs provide a window for inferring the magnetic field and plasma properties at the heliospheric boundary, as well as understanding the complex plasma interactions at the heliospheric boundary. Imaging and global mapping of ENAs refers to a scanning of the global $4\pi$ sky distribution of ENAs flux from the heliospheric boundary. This approach enables a detailed characterization of solar wind suprathermal ions at the heliospheric boundary (specifically downstream of the termination shock), and further aims to reveal the acceleration mechanism of suprathermal ions by turbulence and shocks therein, as well as the multistage charge exchange processes with the interstellar neutral wind \cite{kornbleuth2018globally,baliukin2020heliospheric,baliukin2022energetic,dayeh2022evolution,kornbleuth2023probing}. The observation of ENAs at the heliospheric boundary helps us establish a complete picture of the heliospheric boundary and provides an important panoramic reference for future exploration missions to reach the heliospheric boundary.
\par
So far, several ENA imagers have been launched. The Interstellar Boundary Explorer (IBEX), equipped with two single-pixel cameras (IBEX-Lo and IBEX-Hi), was the first to provide long-term full-sky imaging of ENAs (0.01keV-2keV and 0.5keV-6keV, respectively)\cite{funsten2009ibexhi,fuselier2009ibexlo}. Launched in 2008, IBEX has observed the outer heliosphere for more than 17 years, covering the 24th solar cycle, which has greatly deepened the understanding of the SW--LISM interaction in the outer heliosphere \cite{mccomas2024fourteen}. One of the most remarkable discoveries of the IBEX mission is an unexpected ribbon-like structure with enhanced ENA flux density \cite{funsten2009structures,fuselier2009width,mccomas2009global}. \textbf{The ENA maps from IBEX have provided important constraints on the global structure of the heliosphere. Some interpretations of IBEX observations suggested a more bubble-like heliosphere rather than the classical comet-like configuration \cite{zirnstein2017structure,zhang2024shape}. However, the global shape of the heliosphere remains under active debate, with alternative interpretations involving different degrees of asymmetry and tail structures \cite{schwadron2018heliosphere}}.
The Ion and Neutral Camera (INCA) on board the Cassini spacecraft pieced together a nearly full-sky image of the heliosphere at higher energies (5.2 - 55 keV) from ENA imaging data collected between 2003 and 2009.
Not only did it reveal the power-law distribution of high-energy ENAs but also provided observational evidence for the dynamic processes in the termination shock and heliosheath. It directly demonstrated the importance of high-energy ENAs and and provided a critical observational foundation for quantifying the energy budget of the termination shock and the heliosheath \cite{dialynas2017bubble,krimigis2009imaging,krimigis2010ena}. 
Discoveries by these ENA imagers demonstrate the importance of ENA observations for investigating the heliospheric boundary and motivate the development of next-generation ENA imaging missions. The Interstellar Mapping and Acceleration Probe (IMAP), launched by NASA in 2025 and entering its science phase in February 2026, is expected to provide global ENA maps with improved energy coverage and spatial resolution using its IMAP-Lo, IMAP-Hi, and IMAP-Ultra instruments \cite{mccomas2018interstellar}.
\par
\textbf{
In addition to conventional mechanically collimated ENA imagers, spatial modulation provides an alternative approach for directional imaging, in which the incident angular distribution is encoded into detector measurements through structured absorbers such as grids or coded masks. Unlike direct imaging systems that measure the source distribution through optical focusing or narrow apertures, modulation-based instruments recover the incident source information by solving an inverse problem based on a dedicated forward response model. Therefore, the characterization of the modulation response and the development of appropriate reconstruction algorithms are essential components of such imaging systems.}

\textbf{Spatial modulation techniques have been extensively developed in high-energy solar and astronomical X-ray observations. Although these instruments share the common principle of encoding source information through grid structures, different implementations lead to distinct measurement models and reconstruction strategies. The Reuven Ramaty High Energy Solar Spectroscopic Imager (RHESSI) employed rotating modulation collimators, where the relative motion between front and rear grids during spacecraft rotation produced time-dependent modulation signals. The measured count-rate profiles were subsequently used to recover a limited number of Fourier components (visibilities) of the X-ray source distribution, based on which various reconstruction algorithms, including back projection, CLEAN, maximum entropy, Pixon, and parametric forward fitting, were developed \cite{hurford2002rhessi}. }

\textbf{A different spatial modulation concept is adopted by the Spectrometer/Telescope for Imaging X-rays (STIX) onboard Solar Orbiter and the Hard X-ray Imager (HXI) onboard the Advanced Space-based Solar Observatory (ASO-S). These instruments employ pairs of fixed grids with carefully designed pitch and orientation differences to generate spatial modulation patterns on pixelated detectors. In STIX, each sub-collimator samples a specific Fourier component of the incident X-ray source, and the measured detector counts are related to the corresponding Fourier visibility through a mathematically defined count-distribution model. The image reconstruction problem is therefore formulated as the recovery of the source distribution from a limited number of measured visibilities, requiring dedicated inversion techniques \cite{massa2023stix}. Similar grid-based modulation concepts have been investigated for HXI, where the effects of grid geometry, alignment, and calibration uncertainties on the imaging performance have been systematically evaluated \cite{chen2021design}. These studies demonstrate that, although spatial modulation enables imaging without conventional focusing optics, the achievable reconstruction accuracy depends strongly on the modulation geometry, detector configuration, counting statistics, and the corresponding forward model.}

\par
Conventional heliospheric ENA imagers, such as IBEX-Lo and IBEX-Hi, determine the arrival direction of ENAs primarily through mechanically defined fields of view. Related approaches employing structured apertures have also been developed for planetary neutral-particle imaging. \textbf{For example, the ELENA sensor of the SERENA package onboard BepiColombo uses an oscillating nano-shutter, mechanical gratings, time-of-flight analysis, and a one-dimensional MCP detector array to obtain directional information on neutral particles \cite{Orsini2010SERENA}.} More recently, the Grid-based Energetic Neutral Atom imager (GENA) onboard the Queqiao-2 mission has adopted grid-based spatial encoding combined with time-of-flight measurements for terrestrial ENA imaging \cite{wang2025image, yu2025design}. These developments demonstrate that structured-aperture and grid-based encoding provide an alternative route to directional neutral-particle imaging without relying solely on conventional narrow-field collimation.
Motivated by these developments, the Helios ENA Rapid Mapper (HERMA) has been designed for high-spatiotemporal-resolution imaging of heliospheric ENAs. Each HERMA imaging head employs a stationary spatial-modulation grid in front of a 256-pixel linear silicon detector. ENAs arriving from different directions produce distinct spatial count patterns across the detector because their projected transmission through the grid depends on the incident angle. The incident angular distribution is therefore recovered from the measured detector-count vector through a forward-response model and the corresponding inverse problem. In the present configuration, the reconstruction can be formulated as a non-negative least-squares problem. Under low-count conditions, however, finite-count statistical fluctuations in the detector measurements can lead to appreciable variations in the reconstructed source distribution. We therefore investigate a convolutional neural network (CNN) as an alternative inversion approach and compare its reconstruction performance with that of the conventional optimization method.
\par
In this paper, we propose a reconstructing method for HERMA based on deep learning. We simulate the imaging process based on the observations of IBEX and INCA. The incidence-reception experiment of the HERMA prototype is implemented using Geometry and Tracking 4 (Geant4). Both the least square method and neural network models are used to reconstruct the ENA flux map under different conditions for comparison. We demonstrate the advantages of the neural network model over the traditional algorithm under the condition of low signal-to-noise ratio. In Section 2, we introduce the HERMA mission and describe the experimental setup of simulation, including the inversion algorithm for reconstructing the ENA flux distribution in the source region. In Section 3, we present the simulation results at different staring times and incident angles, and we compare the performance of different inversion algorithms in reconstructing the incident particle source. In Section 4, we summarize this study and discuss improvements to be carried out in the future.
\section{Materials and Methods}
\subsection{HERMA Instrument Configuration and Mission Profile}
\begin{figure}[h]
    \centering
    \includegraphics[width=0.95\textwidth]{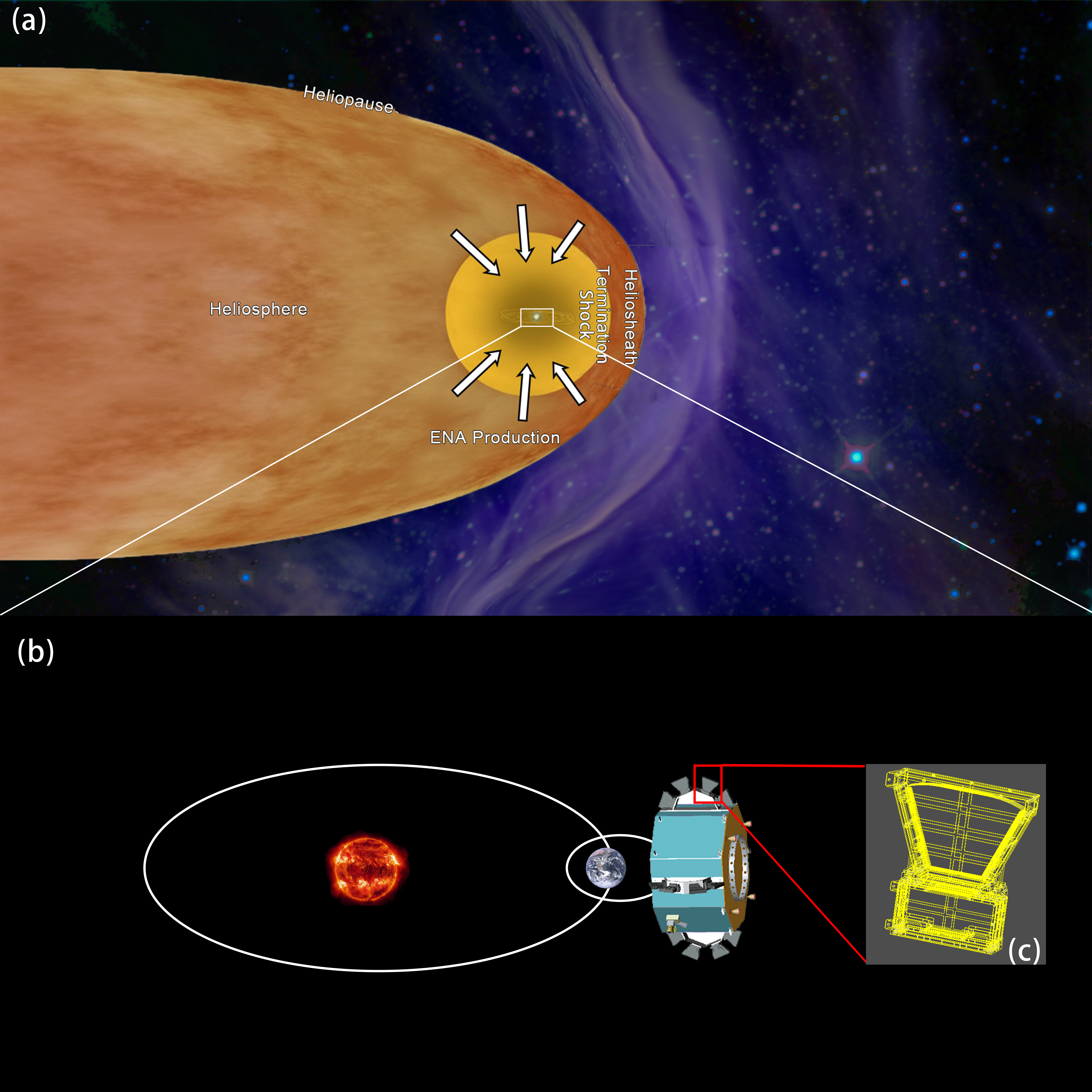}
    \caption{Overview of the orbit and configuration of the HERMA instrument. (a)Schematic illustration of the heliosphere and ENA production region, adapted from an artist’s depiction by NASA’s Goddard Space Flight Center Conceptual Image Lab; (b) Schematic diagram of the HERMA orbit, the $4\times4$ arrangement of imaging detectors. (c)The structure of a single imaging detector.}
    \label{fig:1}
\end{figure}

\textbf{HERMA is a spatial-modulation ENA imager designed for rapid mapping of heliospheric ENAs over a target energy range of 10–500 keV. A detailed description of the instrument design, imaging principle, and ground calibration is provided by Chen et al. (2026)\cite{chen2026ground}. Here, we briefly summarize the instrument configuration and describe the mission profile relevant to the present forward-modeling and inversion study.
This payload consists of 16 identical imaging heads divided into four groups, which are mounted on four side panels oriented $90^\circ$ apart around the spacecraft body (Fig.\ref{fig:1}(b)). 
Each head has a fan-shaped field of view (FOV) of $45^\circ \times 2.5^\circ$ and a geometric factor of about $0.22~\mathrm{cm^2\,sr}$. 
Within each group, four coplanar imaging heads are arranged side by side, such that their individual FOVs form a combined fan-shaped instantaneous FOV of $180^\circ \times 2.5^\circ$.} The four heads within one group are arranged with no overlap between adjacent FOVs.

\textbf{
Each imaging head (Fig.\ref{fig:1}(c)) comprises a fan-shaped passive collimator, a charged-particle deflection section, a spatial-modulation grid, and a 256-pixel linear silicon detector. 
The collimator limits the incident direction. 
Electrodes located on the two sides of the collimator produce an electric field that deflects charged particles. 
The spatial-modulation grid is made of tungsten and has an overall size of $127.0~\mathrm{mm}\times35.0~\mathrm{mm}$, with a coded region of $114.5~\mathrm{mm}\times25.0~\mathrm{mm}$. The coding pattern contains 72 transmitting gaps with widths ranging from 0.45 to 3.60 mm. These gaps are distributed over two sets of spatial Fourier periods, following the modulation concept described by Hurford et al. \cite{hurford2002rhessi}; the detailed coding geometry and ground calibration are provided by Chen et al.\cite{chen2026ground}. The use of multiple gap widths introduces multiple spatial scales into the transmission pattern. As the incident direction changes, the projected positions and lengths of the transmitting gaps on the silicon detector also change, producing direction-dependent count distributions across the 256 detector pixels. These modulated count patterns encode the angular information of the incident ENA distribution.
The detector is based on an ultra-thin-dead-layer design to support the targeted 10-500 keV energy range; the design energy resolution is $\leq 10\%$ at 100 keV.}

\textbf{
The HERMA mission profile uses a highly elliptical Earth orbit with a nominal perigee altitude of approximately 48,000 km, an apogee altitude of approximately 320,000 km, and an inclination of $28.5^\circ$. The high apogee provides long observing arcs outside the near-Earth magnetic environment, while an approximately 3:1 resonance with the lunar orbital period is used in the baseline design to limit the cumulative effect of lunar perturbations. 
In survey mode, the spacecraft face carrying the body-mounted solar array remains pointed toward the Sun, while the four payload groups mounted on the side panels form four fan-shaped fields of view, each covering $180^\circ \times 2.5^\circ$.}

\textbf{The three-axis-stabilized spacecraft steps about the Sun-pointing axis through 36 increments of $2.5^\circ$, covering a total rotation of $90^\circ$. After each $2.5^\circ$ maneuver, the spacecraft holds its new orientation fixed in inertial space for a staring observation, during which the detectors accumulate ENA counts. 
Combining the four orthogonal fan-shaped sectors with the stepwise scan provides nearly $4\pi$ sky coverage.}

\subsection{Forward Response Model}
\textbf{
During a single staring observation, the azimuthal field of view remains fixed with a width of $2.5^\circ$, while the instrument measures the ENA flux over a $45^\circ$ range in elevation. ENAs arriving from different directions within this field of view are simultaneously modulated by the grid, and their combined contributions produce a count distribution, $\boldsymbol{D}$, across the 256 detector elements.}
\textbf{
To describe this imaging process, we define an ideal response function $f$, which maps an incident ENA distribution $\boldsymbol{S}$ to the mathematically expected detector response $\boldsymbol{\bar{D}}$:}

\begin{equation}
    \boldsymbol{\bar{D}}=f(\boldsymbol{S}).
\end{equation}
\textbf{We approximate the continuous full-sky ENA flux distribution by a two-dimensional matrix of angular pixels. For a single imaging head at a fixed azimuthal position, its $45^\circ \times 2.5^\circ$ field of view corresponds to a one-pixel-wide strip of this full-sky map. Along the elevation direction, this strip is divided into 19 angular bins centered at $\theta_j$, spanning from $-22.5^\circ$ to $22.5^\circ$ with an angular interval of $\Delta\theta=2.5^\circ$. The ENA flux values in these 19 bins constitute the source vector $\boldsymbol{S}$.}
\textbf{
For the forward-response calculation, the ENAs within each angular bin $j$ are approximated as a parallel beam incident at the central angle $\theta_j$. Under this angular discretization and the parallel-beam approximation, the response function $f$ is represented by a linear operator. Because the detector response is the superposition of the contributions from all angular bins, the ideal detector response can be expressed as the following linear forward model:}

\begin{equation}
\boldsymbol{\bar{D}}=\mathbf{M}\cdot\boldsymbol{S},
\end{equation}

\textbf{
where $M_{ij}$, the $(i,j)$th element of $\mathbf{M}$, represents the normalized response of detector strip $i$ to an ENA beam incident from angular bin $j$. $M_{ij}$ can be estimated geometrically as:}

\begin{equation}
    M_{ij}=\frac{l_{ij}\cos\theta_j}{w_{\mathrm{strip}}},
\end{equation}

\textbf{
where $l_{ij}$ is the length of the grid’s open aperture projected onto detector strip $i$ for an incident angle $\theta_j$, and $w_{\mathrm{strip}}$ is the width of the detector strip used for normalization.}

\par
\begin{figure}[h]
    \centering
    \includegraphics[width=0.95\textwidth]{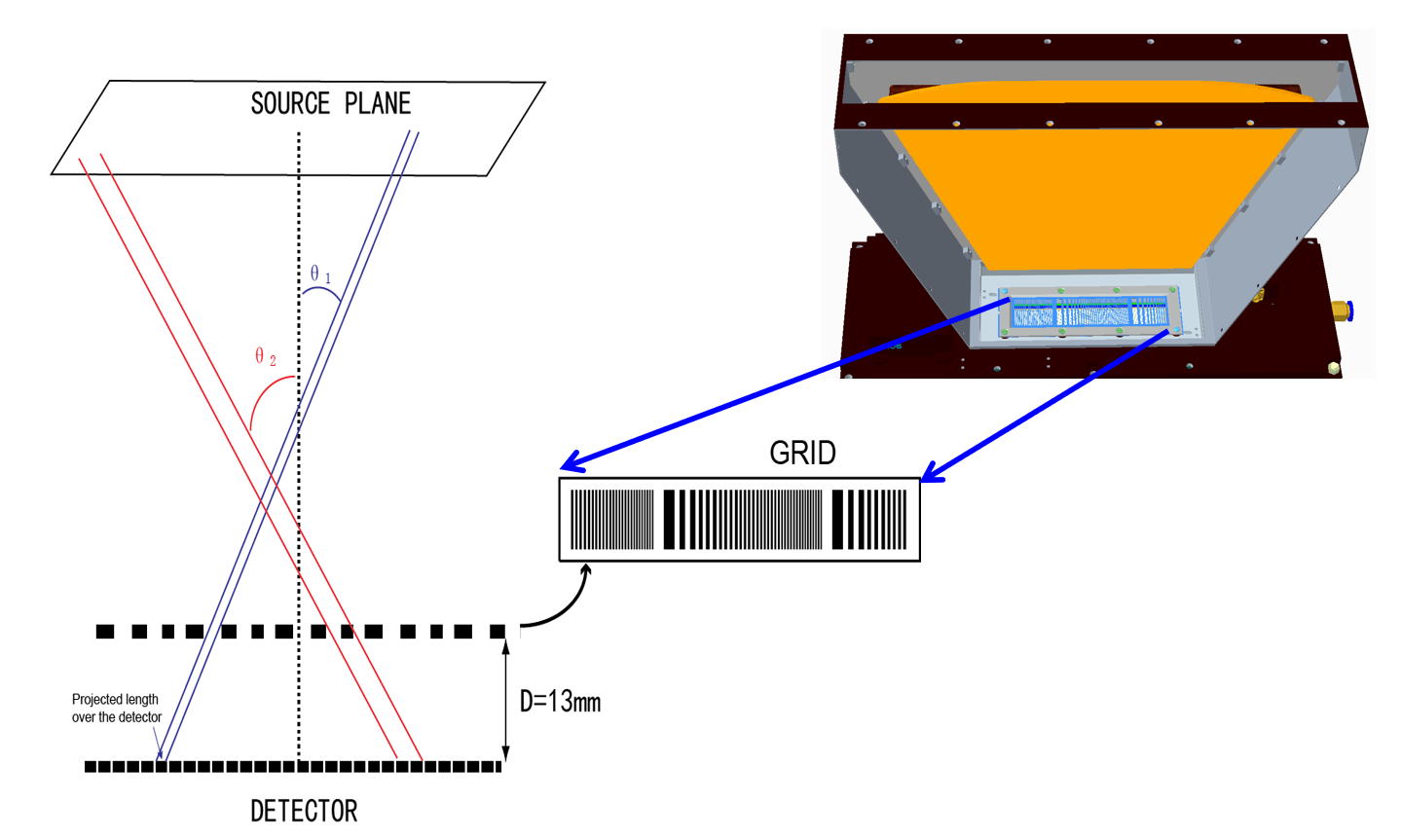}
    \caption{Right: Schematic diagram of the structure of a single HERMA imaging detector (including deflection plates, a modulation grid and a silicon semiconductor detection unit with 256 pixels). Left: Incident particles modulated by the grid. The red and blue lines denote the particles incident from the source plane at incident angles of $\theta_1$ and $\theta_2$, respectively. $D=13\mathrm{mm}$ represents the distance from the grid to the sensors.}
    \label{fig:2}
\end{figure}
\subsection{Geant4 Simulation Settings}
\par
Geant4 is a Monte Carlo simulation toolkit developed by European Organization for Nuclear Research (CERN), it can be used to model and simulate the interactions between particles and matter across a wide range of energy scales, from low-energy atomic processes to high-energy particle physics experiments\cite{siegl2009standard}.
With the help of Geant4, we carried out simulation experiments on the process of ENAs hitting the detector. In Geant4, we reconstructed the detailed structure of the detector, which consists of 72 solid iron grid cells and 72 hollow grid cells. A rectangular source with dimensions $115.2\text{mm}\times20\text{mm}$ was placed at a distance $200$mm from the detector. \textbf{The rectangular source plane is a numerical injection surface and does not represent the physical size or shape of the heliospheric ENA source region. Because the heliospheric ENAs source is remote compared with the characteristic dimensions of the instrument, ENAs arriving from a given angular bin are approximated as a locally parallel beam. The source plane is chosen to cover the active detector area so that the response of the instrument to each incident direction can be evaluated.} For a rectangular source with an area $\Delta A$, normal vector $\vec{n}$ and incident direction $\vec{i}$, the number of particles emitted by the rectangular source within the staring time $\Delta t$ can be calculated by the formula:
\begin{equation}
    N=J\Delta A\Delta E\Delta \Omega\Delta t\cos\langle\vec{n},\vec{i}\rangle
\end{equation}
where $J$ is the differential flux of the particle source, $\Delta E$ is the energy bin width, $\langle\vec{n},\vec{i}\rangle$ represents the angle between $\vec{n}$ and $\vec{i}$, and $\Delta \Omega$ is the solid angle width.
We carried out hydrogen ENA incidence simulation experiments with $\Delta t=2000s, 5000s,10000s,20000s$, respectively. The incident angle range of particles is from $-22.5\degree$ to $22.5\degree$ with an angular step of $2.5\degree$. The differential flux $J$ of the incident particle source is derived by interpolating INCA-observed ENA flux data onto a $2.5\degree\times2.5\degree$ grid. The data cover the energy range $5.2\text{keV}\sim 13.5\text{keV}$, latitude $-90\degree\sim-45\degree$, longitude $0\degree\sim 90\degree$, and span the period 2003--2009. This corresponds to the field of view imaged by a single detector of the HERMA detector groups when all 16 detectors work together for the full-sky imaging.
\subsection{Inversion Algorithms}
\par
Following the response of the silicon detector, $\boldsymbol{D}$ in Geant4, we perform inversion and reconstruct the incident source $\boldsymbol{S}$, i.e., the angular flux distribution of the incident ENAs. The reconstruction result serves to assess the ability of the HERMA detector to resolve the incident direction and intensity of ENAs. In this study, we test the performance of two inversion methods, respectively. The first method is the Sequential Least Squares Quadratic Programming (SLSQP) algorithm, which is designed to solve the following optimization problem:
\begin{equation}
\begin{split}
      &\mathrm{argmin}_{x\in \mathbb{R}^n}f(x)\\
    \mathrm{s.t.}\ &c_i(x)=0\ \ \ (i=1,2,\cdots,p)\\
    &c_i(x)\le0\ \ \ (i=p+1,\cdots,p+q),
\end{split}
\end{equation}

\textbf{where $f(x)$ is the objective function, and $c_i(x)$ is the $i$-th constraint. $p$ denotes the number of equality constraints, while $q$ denotes the number of inequality constraints. In our case, the objective function is defined as $f(S) = \|\mathbf{D}-\mathbf{M}\cdot \boldsymbol{S}\|_2^2$, subject to the constraint $S_j\geq0$, where the subscript $2$ denotes the Euclidean($L_2$) norm and the superscript $2$ indicates that this norm is squared.}
The core idea of the SLSQP algorithm is to approximate the optimal solution to the original nonlinear optimization problem by iteratively solving a sequence of quadratic programming (QP) subproblems. The least-squares technique is used to efficiently solve these subproblems, thereby avoiding the complexity of directly solving quadratic programming problems\cite{kraft1988software}. 

\par
\begin{figure}[h]
    \centering
    \includegraphics[width=0.95\textwidth]{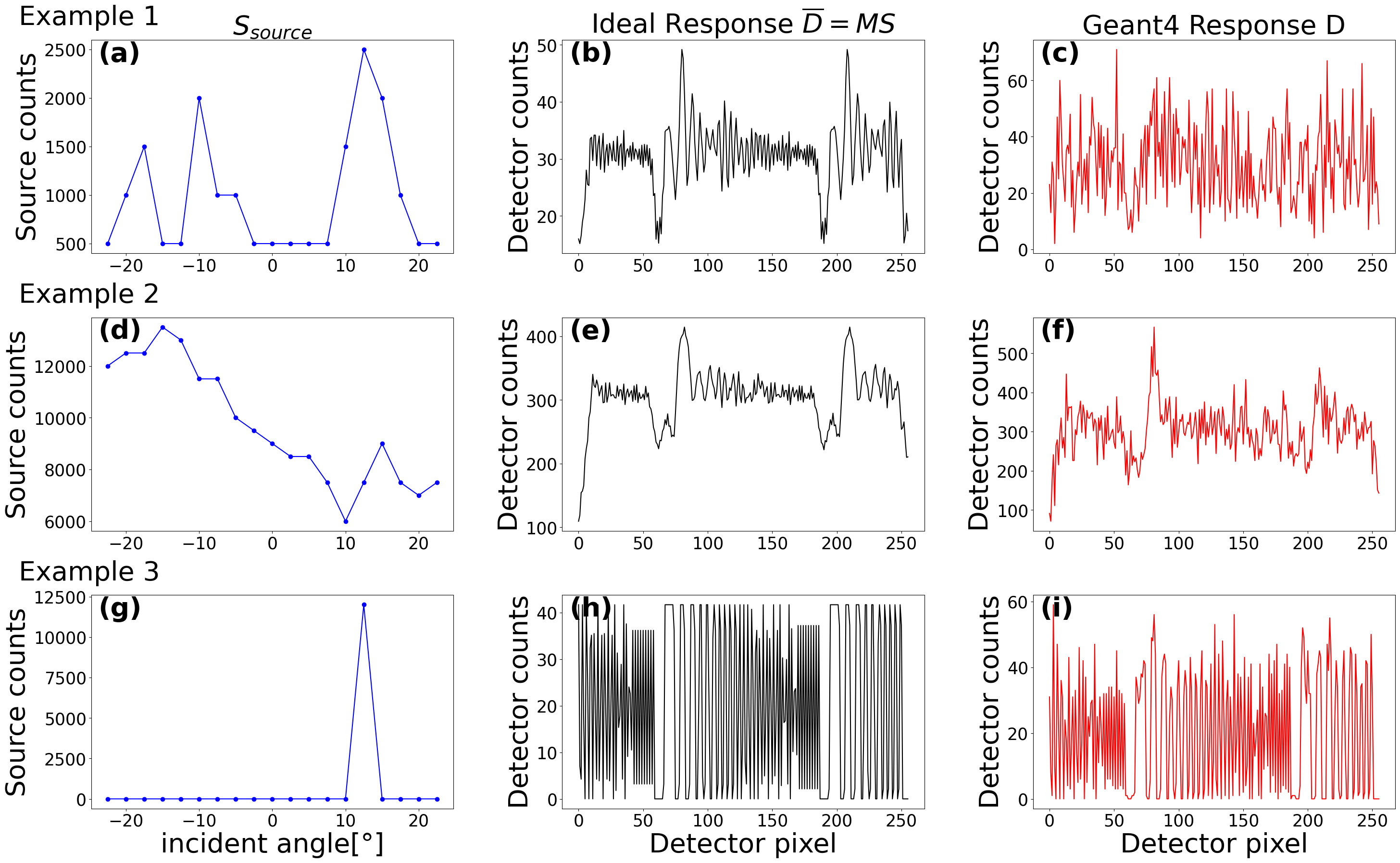}
    \caption{Examples of the CNN training samples generated from Geant4 simulations. For each example, the left column shows the incident ENA source distribution $\boldsymbol{S}$, the middle column presents the ideal detector response $\bar{D}=MS$ calculated from the response matrix, and the right column shows the detector counts $D$ obtained from Monte Carlo particle transport in Geant4.}
    \label{fig:3}
\end{figure}
\begin{figure}[h]
    \centering
    \includegraphics[width=0.95\textwidth]{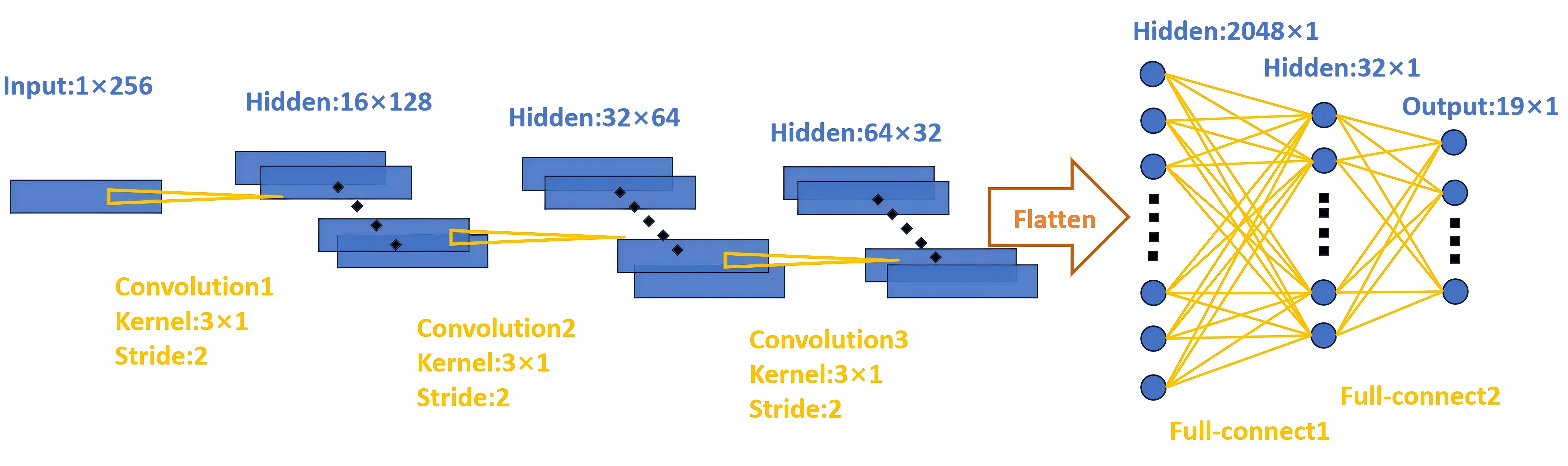}
    \caption{Framework of the CNN deep learning algorithm adopted in this study. Dark blue icons and golden icons represent data and data operations, respectively. The text in the same color as the data or operations provides the corresponding explanatory information. The rectified linear unit (ReLU) is used as the activation function for all layers except the last layer that adopts the Softplus function.}
    \label{fig:4}
\end{figure}

The second method is to use a convolutional neural network (CNN) as a solver of the inversion problem. The reason to introduce this method is that SLSQP performs poorly in extreme cases. When the incident ENA flux is low and the staring time is short, the detected response of the silicon detector, $\boldsymbol{D}$ often exhibits significant deviations from the expected response $\boldsymbol{\bar{D}}$ due to finite-count statistical fluctuations, which substantially degrades the performance of the conventional SLSQP algorithm. In the field of deep learning, CNN is widely used for feature extraction of structured data. Here, CNN can learn the distribution of real  response $\boldsymbol{D}$ under different incident source $\boldsymbol{S}$ to improve the inversion reliability. The model structure we use is shown in Fig. \ref{fig:4}. The input is a $1\times256$ 1D vector of real response $\boldsymbol{D}$, while the output is a $19\times1$ vector of corresponding incident source, $\boldsymbol{S}$. The model has 3 convolutional layers, 1 fully connected layer and 1 output layer. The activation function of each layer adopts the Rectified Linear Unit (ReLU: $\max\{0,x\}$), except for the last layer using the Softplus function $\ln(1+e^x)$.
\textbf{During the training process, we generated 2000 source–response pairs using Geant4 simulations. The incident source distributions S were constructed from several representative angular morphologies, including narrow single-peak, multi-peak, broad, and smoothly varying profiles. The parameters characterizing these profiles, such as the overall intensity, peak position, peak amplitude, and profile width, were randomly varied to provide sufficient diversity in the training set. These profiles are designed to cover possible angular structures of heliospheric ENA signals and improve the robustness of the inversion model. The resulting training pair consists of the source vector S and the detector count vector $D$. Figure \ref{fig:3} presents several examples of these training samples.
The source distributions were discretized over 19 incident directions spanning $-22.5\degree$
 to $22.5\degree$ with an angular interval of $2.5\degree$.}
The dataset was then split into training and validation sets at an 8:2 ratio. \textbf{The network was trained using the Adam optimizer with a fixed learning rate of $5\times10^{-4}$. The mean squared error (MSE) between the reconstructed and target source vectors was used as the loss function. A batch size of 20 was adopted, and the network was trained for 400 epochs.} After the training is complete, we set as the test set the Geant4 simulation results with the ENA sky-map data of $5.2 \text{keV}\sim13.5 \text{keV}$ detected by INCA from 2003 to 2009 as the input, but varying the total incident number of particles (corresponding to different staring times), to test the inversion performance of the model under different conditions.

\section{Results and Discussion}
\subsection{Geant4 Simulation Results: the effect of \textbf{varying} the staring time}
\par
\begin{figure}[h]
    \centering
    \includegraphics[width=0.95\textwidth]{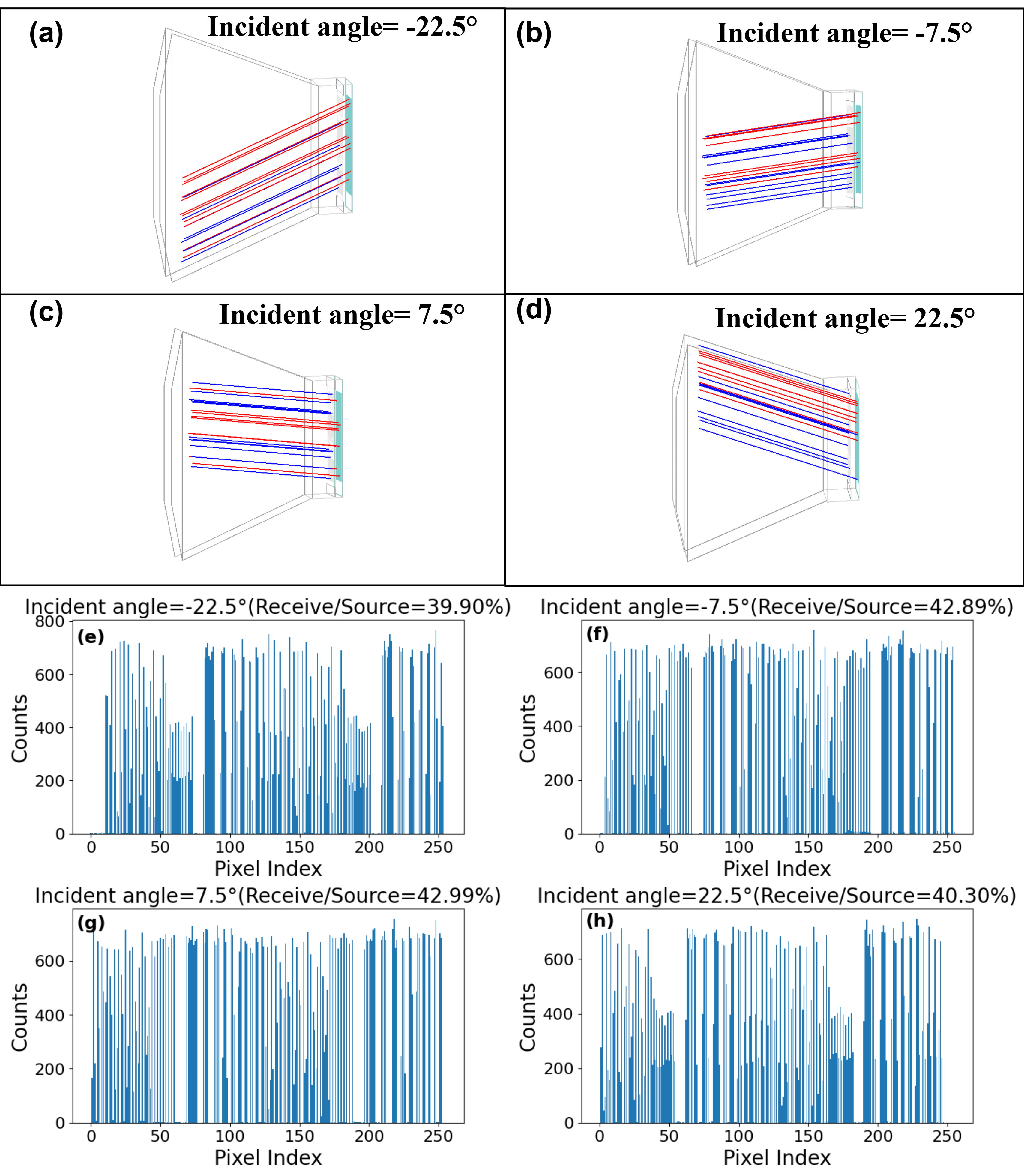}
    \caption{(a)--(d) ENA particle trajectories incident on the HERMA detector, simulated using Geant4 at incident angles of $-22.5\degree$, $-7.5\degree$, $7.5\degree$, and $22.5\degree$, respectively. The red lines represent particles collected by the detector, while the blue lines represent those blocked by the grid. (e)--(h) ENA counts of each pixel in the detector  at incident angles of $-22.5\degree$, $-7.5\degree$, $7.5\degree$, and $22.5\degree$, respectively.}
    \label{fig:5}
\end{figure}

Fig. \ref{fig:5}(a)--(d) show the trajectories of ENAs coming into the HERMA detector simulated by Geant4 when the incident angles are $-22.5\degree$, $-7.5\degree$, $7.5\degree$ and $22.5\degree$, respectively. The detector can receive particles with incident angles from $-22.5\degree$ to $22.5\degree$ and modulate the incident source periodically through the grid. About 40\% ENAs will be received by the detector, while others will be blocked by the grid.
There will be different response distributions in the detector with the variation of the incident angle (Fig. \ref{fig:5}(e)--(h)). These distribution patterns restore the information of the incident ENAs, which can be reconstructed by inversion.

\par
\begin{figure}[h]
    \centering
    \includegraphics[width=\textwidth]{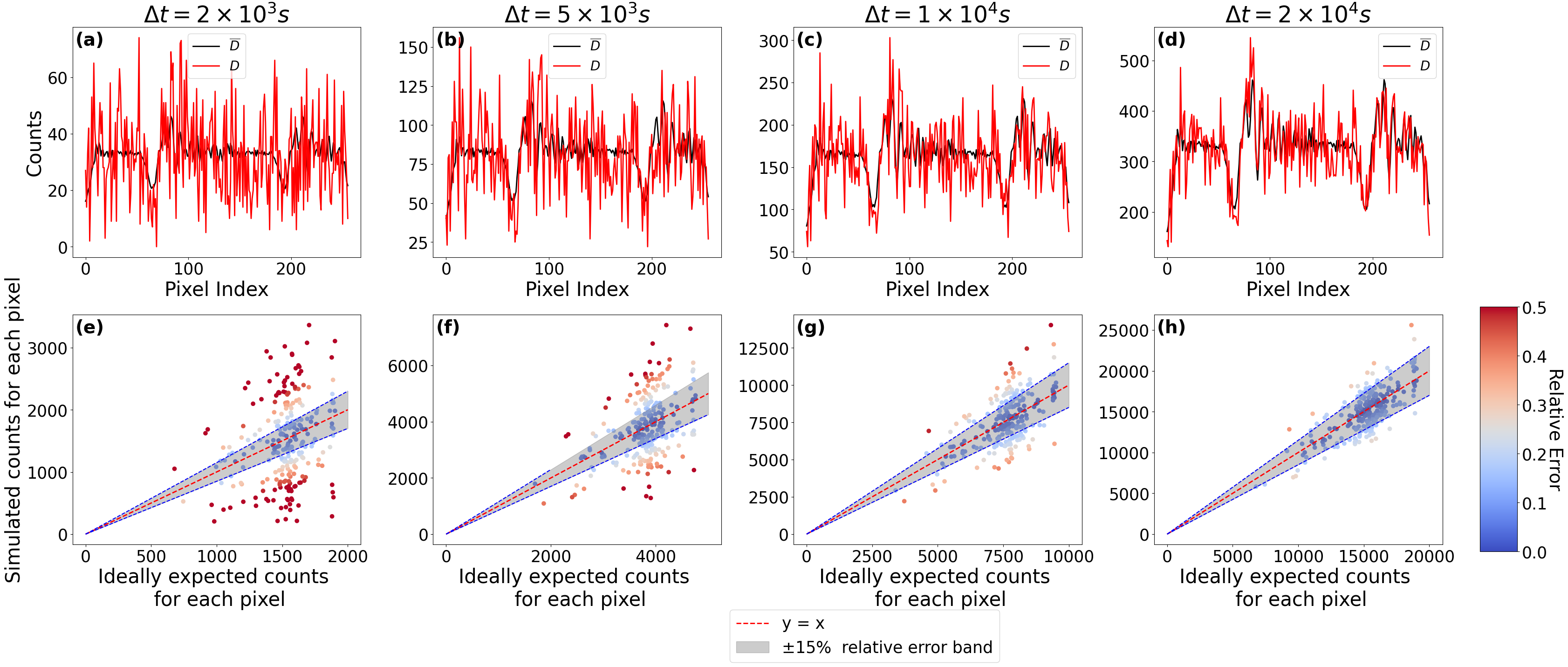}
    \caption{The upper panels (a--d) show the comparison between the ideal detector response 
$\boldsymbol{\bar{D}}$(black curves) and the simulated detector counts $\boldsymbol{D}$ (red curves) for three different staring times,$\Delta t=2\times 10^3\rm{s}$,$\Delta t=5\times10
^3\rm{s}$, $1\times10^4\ \rm{s}$, and $2\times 10^4\ \rm{s}$, respectively.Scatter plot of the detector pixel responses obtained from Geant4 simulations versus the ideally expected particle responses of each pixel calculated using the projection matrix at a staring time of (e)$\Delta t=2\times 10^3\rm{s}$;  (f) $5\times10^3$ s; (g) $1\times10^4$s; (h) $2\times10^4$s. The red dashed line represents $y=x$, and the shaded region indicates the range where the deviation between the simulated data and the ideally expected results is less than 15\%.}
    \label{fig:6}
\end{figure}

\textbf{Figure \ref{fig:6} compares the detector responses obtained from Geant4 simulations with the ideal responses calculated from the forward model. The upper panels (a–-d) show representative detector count profiles for different staring times ($\Delta t=2\times 10^3\rm{s}$,$\Delta t=5\times 10^3\rm{s}$,$1\times10^4\rm{s}$, and $2\times10^4\rm{s}$). The Geant4 results fluctuate around the ideal responses due to the finite number of particles detected in the Monte Carlo simulation.}
\textbf{
The lower panels Fig. \ref{fig:6}(e--h) show the corresponding pixel-by-pixel comparison between $\boldsymbol{D}$ and $\boldsymbol{\bar{D}}$. Each point represents one detector pixel, and the dashed line indicates the ideal agreement ($y=x$). The deviation from the diagonal line decreases with increasing staring time because more particles are accumulated and the statistical fluctuation is reduced. The color of each point indicates the relative deviation between the simulated and ideal counts. For a staring time of $2\times10^4\rm{s}$, most detector pixels exhibit relative deviations within 15\% (Figure \ref{fig:6}(h)), demonstrating that the Geant4 simulations reproduce the expected detector response with finite-count statistical fluctuations. In particular, the 2000s case exhibits the strongest deviations (Figure \ref{fig:6}(e)), indicating that finite-count statistical fluctuations become increasingly important in this low-count regime.}
\begin{figure}[h]
    \centering
    \includegraphics[width=0.95\textwidth]{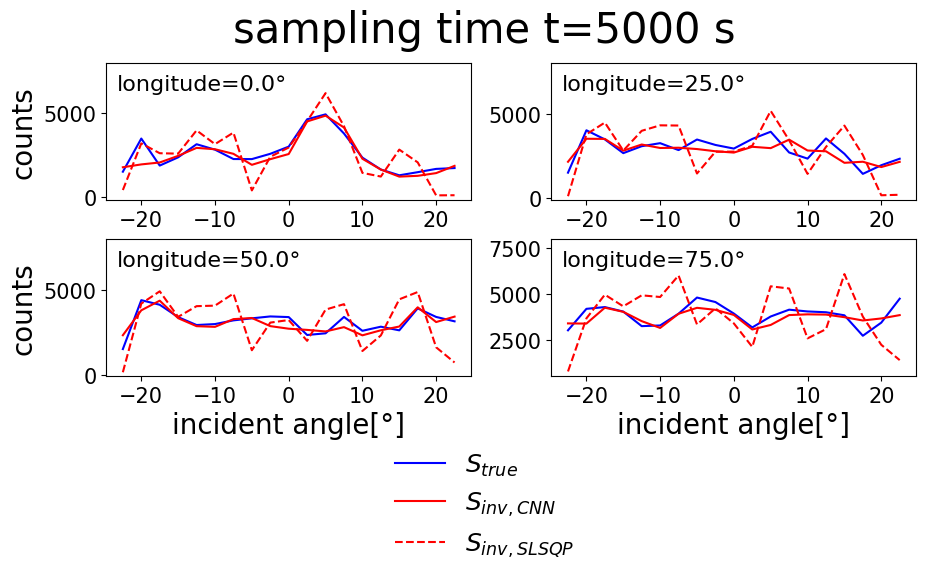}
    \caption{Comparison of reconstructed ENA source distributions obtained by CNN and SLSQP inversion at different longitudes.The reconstructed source angular distributions are shown at four representative longitudes: (a) $0^\circ$, (b) $25^\circ$, (c) $50^\circ$, and (d) $75^\circ$, for a staring time of $\Delta t=5000\rm{s}$. The blue curves represent the original source distributions $\boldsymbol{S}_{\rm{true}}$, while the solid red and dashed red curves denote the reconstructed distributions from the CNN model ($\boldsymbol{S}_{\rm{inv,CNN}}$) and the SLSQP optimization method ($\boldsymbol{S}_{\rm{inv,SLSQP}}$), respectively. The source distributions are reconstructed independently from the corresponding simulated HERMA detector measurements. }
    \label{fig:7}
\end{figure}
\subsection{Performance Comparison of Inversion Algorithms}
\textbf{
To evaluate the performance of the proposed CNN inversion method, we compared it with the conventional SLSQP optimization method. Figure \ref{fig:7} presents representative inversion results for four different longitudes with a staring time of 5000 s. The blue curves represent the true source distributions, while the red solid and red dashed curves denote the reconstructed sources from CNN and SLSQP, respectively.}

\textbf{Both methods are able to recover the main structures of the source distribution. However, the SLSQP inversion exhibits larger deviations, especially for regions with weak signals. This is mainly because the inversion problem is ill-posed and sensitive to statistical fluctuations in the detector measurements. In contrast, the CNN model, trained using Geant4-generated detector responses, provides more stable reconstruction results by learning the nonlinear mapping between detector counts and source distributions.}
\par
\begin{figure}[h]
    \centering
    \includegraphics[width=0.95\textwidth]{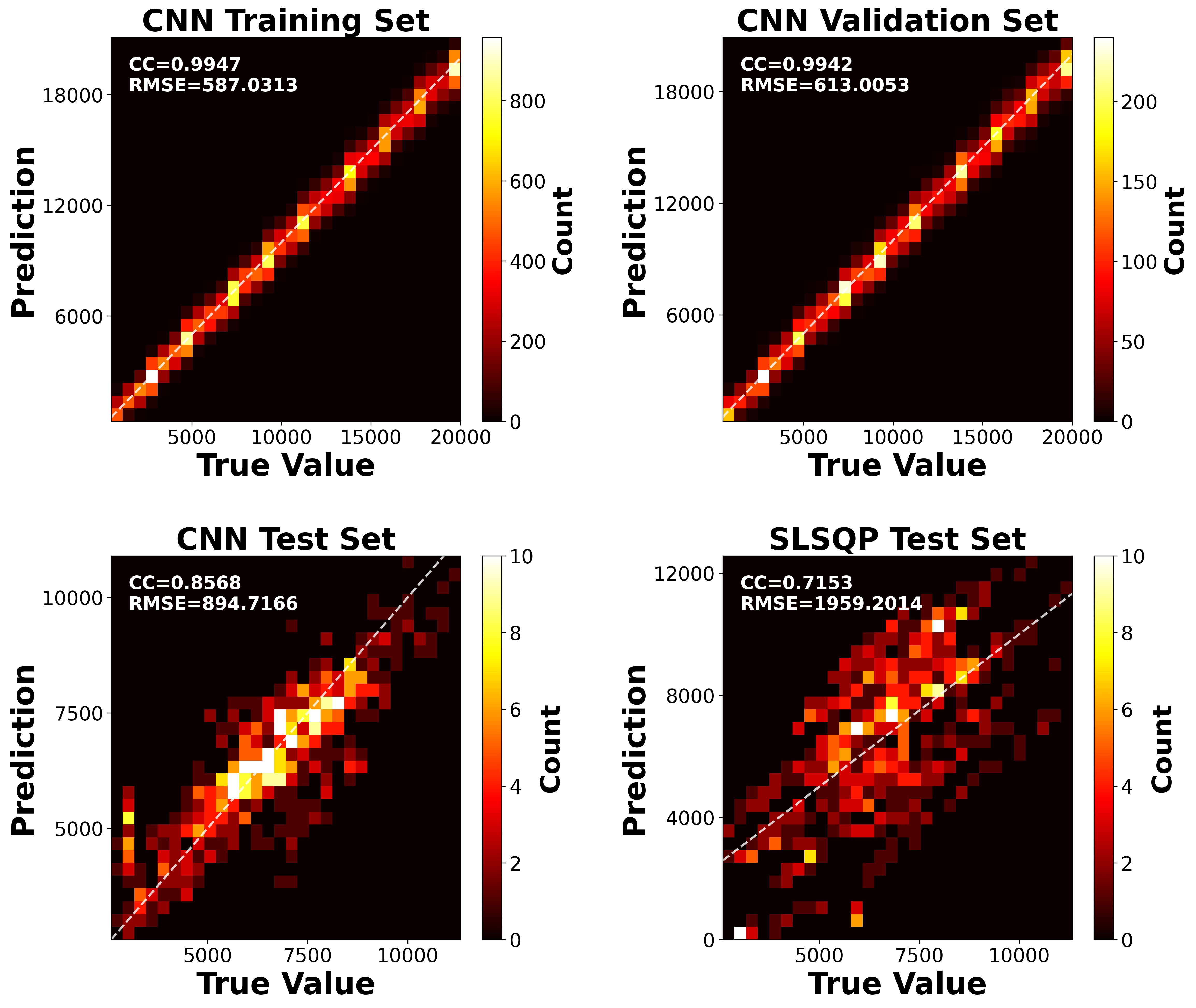}
    \caption{(a) Comparison of ideally expected incident source particle counts versus CNN-inverted incident particle counts on (a: the training set), (b: the validation set), (c: the test set). (d) Comparison of ideally expected incident particle counts versus incident particle counts inverted by the SLSQP algorithm on the test set. The white dashed line denotes the line $y=x$.}
    \label{fig:8}
\end{figure}

We make a further comparison between the performance of SLSQP algorithm and CNN method. The performance metrics are RMSE and Pearson CC. They are defined as,

\begin{equation}
\text{RMSE}(\boldsymbol{x},\boldsymbol{y}) = \sqrt{\frac{1}{n} \sum_{i=1}^{n} (x_i - y_i)^2}
\end{equation}
\begin{equation}
    \text{CC}(\boldsymbol{x},\boldsymbol{y}) = \frac{\sum_{i=1}^{n} (x_i - \bar{x})(y_i - \bar{y})}{\sqrt{\sum_{i=1}^{n} (x_i - \bar{x})^2} \sqrt{\sum_{i=1}^{n} (y_i - \bar{y})^2}}    
\end{equation}
where $\boldsymbol{x}$ and $\boldsymbol{y}$ are n‐dimensional vectors, and $\bar{x}$ is the arithmetic average of $\boldsymbol{x}$. Therefore, $\text{RMSE}(\boldsymbol{S}_{\text{ture}},\boldsymbol{S}_{\text{pred}})$ represents the general deviation of the prediction, and $\text{CC}(\boldsymbol{S}_{\text{ture}},\boldsymbol{S}_{\text{pred}})$ represents the ability of our inversion methods to predict the
source counts for each incident angle, where $\boldsymbol{S}_{\text{ture}}$ and $\boldsymbol{S}_{\text{pred}}$ stand for the true value and predicted value of the incident source array $\boldsymbol{S}$.
Fig. \ref{fig:8}(a) and (b) show the performance of the CNN model on the training set and the validation set, respectively. The results demonstrate that the trained network yields accurate inversion results for both the training and validation sets. 
At the same time, we compared the performance of the CNN inversion method and the SLSQP inversion algorithm on the test set (Fig.~\ref{fig:8}(c) and (d)). The inversion results of the CNN method have a higher correlation coefficient and a smaller RMSE than the SLSQP algorithm.
\par
\begin{figure}[h]
    \centering
    \includegraphics[width=\textwidth]{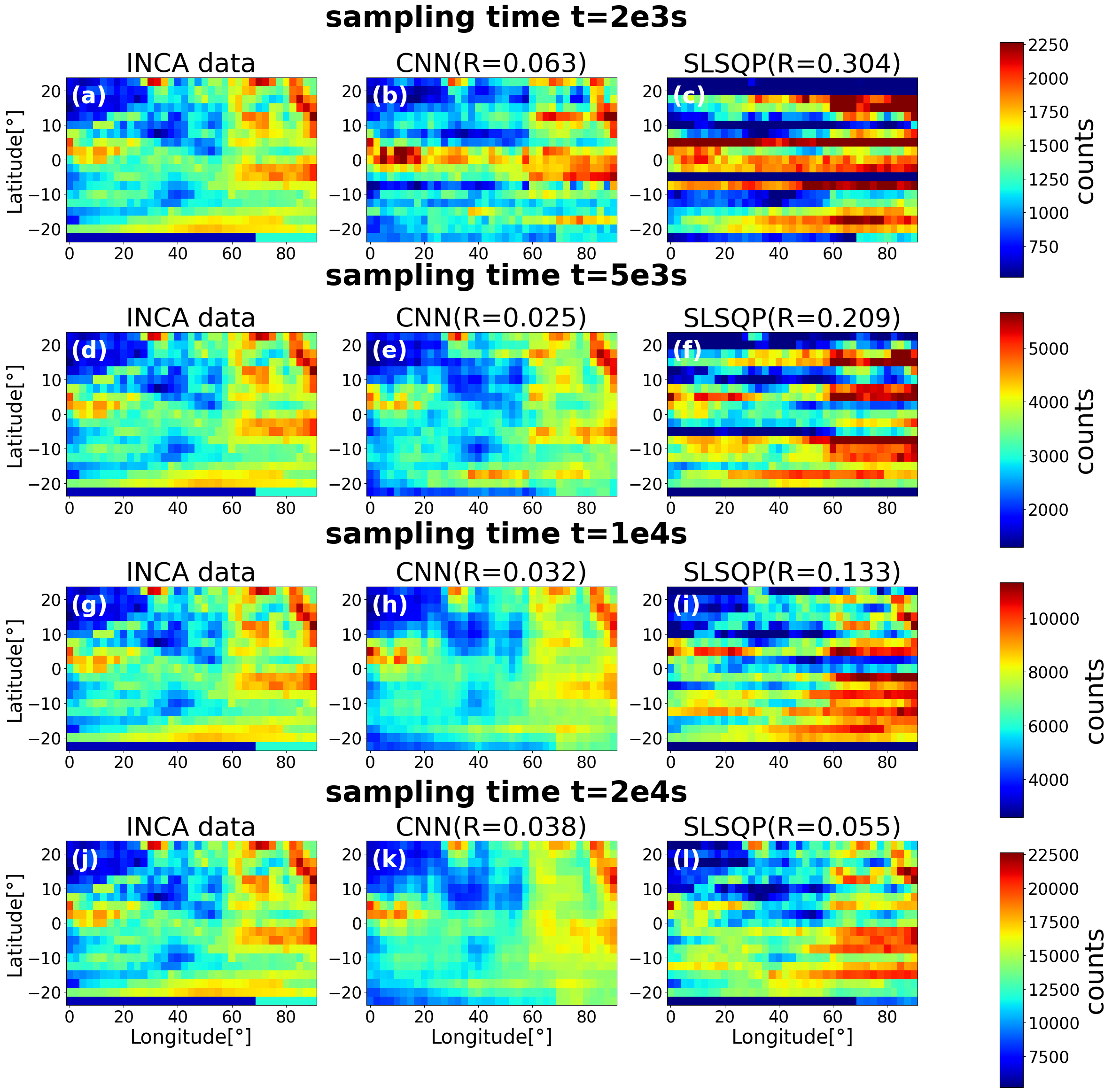}
    \caption{(a) Interpolation results of ENA incident particle number distribution data detected by INCA over latitudes from $-90\degree$ to $-45\degree$ and longitudes from $0\degree$ to $90\degree$ on a $2.5\degree\times2.5\degree$ grid at a staring time of 5000 s.(b) Incident ENA flux distribution reconstructed by the CNN model.(c) Incident ENA flux distribution reconstructed by the SLSQP algorithm.(d)-(f) Results at a staring time of 10000 s.(g)-(i) Results at a staring time of 20000 s.}
    \label{fig:9}
\end{figure}

Fig. \ref{fig:9} shows the spatial distribution of the incident source $\boldsymbol{S}$ obtained by inverting the detector response with the SLSQP algorithm and the CNN method when the staring times are $\Delta t=2000s, 5000s,10000s,20000s$, respectively. Under each staring time, the incident source distribution reconstructed by the CNN inversion method (Fig. \ref{fig:9}(b), (e), (h)) is very close to the ideally expected incident source distribution shown in Fig. \ref{fig:9}(a), (d), (g). However, when the staring times are $\Delta t=5000s,10000s$, the incident source distribution reconstructed by the SLSQP algorithm has a significant deviation from the ideally expected incident source distribution (Fig. \ref{fig:9}(c), (f)), which is due to the large difference between the simulated detector response $\boldsymbol{{D}}$ and the expected detector response $\boldsymbol{\bar{D}}$ . For example, the SLSQP algorithm cannot reproduce the high-flux region at $(60\degree\sim80\degree,0\degree)$ very well.
We use a parameter $R$ to further quantify the reconstruction performance of different inversion methods.
\begin{equation}
    R= \frac{\sum_{i=1}^{N_{\theta}}\sum_{j=1}^{N_\phi}(\frac{J_{\text{INCA}}-J_{\text{inv}}}{J_{\text{INCA}}})^2}{N_\theta N_\phi}
\end{equation}
where $N_\theta$ and $N_\phi$ are the number of pixels in the elevation direction and angular direction respectively, $J_{\text{INCA}}$ and $J_{\text{inv}}$ represent the input INCA data and the inverted ENA flux data, respectively.
A smaller value of the metric $R$ corresponds to a better inversion result. It can be seen from Fig. \ref{fig:9} that when the staring times are $\Delta t=2000s,5000s,10000s,20000s$, the $R$ values obtained by the CNN inversion algorithm are 0.063, 0.025, 0.032 and 0.038, respectively, which are all better than the results obtained by the SLSQP algorithm. In addition, the incident source distribution results reconstructed by the CNN inversion algorithm have lower dependence on the staring time within the range $5\times10^3 \text{s}\sim 2\times10^4\text{s}$, showing higher accuracy and stability than the traditional SLSQP algorithm. At the shortest staring time of 2000 s, both inversion methods are more strongly affected by finite-count statistical fluctuations. The degradation is particularly evident in the SLSQP reconstruction, where pronounced band-like structures appear. The CNN reconstruction retains the major large-scale features of the input distribution, although its deviation also increases compared with the 5000 s case. Based on this, we prove that a single detector of the HERMA, equipped with the CNN inversion model, is capable of accurate imaging of a celestial sphere range with a longitude of $90\degree$ and a latitude of $45\degree$ within about $10000 \mathrm{\ s}\times 36=360000 \mathrm{\ s}$ (i.e., about four days). Thus, when the 16 detectors of HERMA work together, it can perform high-time-resolution imaging of ENAs in the energy range of $5.2 \text{keV}\sim13.5 \text{keV}$ within about 4 days, which represents a significant advancement over previous ENA exploration missions.

\section{Conclusion}
\par
This study introduces a simulation experiment of the HERMA imaging instrument carried out with Geant4 and different inversion methods. In the experiment, we simulated the imaging process of HERMA for different staring durations using INCA observations as the incident flux for reference, and compared the reconstructed incident particle source $\boldsymbol{S}$ obtained from the detector response with different methods. The CNN inversion method is proved to provide higher reconstruction accuracy and stability than the traditional inversion SLSQP algorithm under short staring times. This approach can
effectively enable high-resolution, rapid ENA imaging for the HERMA instrument, and thus provides core algorithmic support for the instrument's engineering verification and on-orbit application, bridging the gap in wide-energy-range, high-resolution ENA detection, and contributes to a deeper understanding of the interaction mechanisms between the solar wind and interstellar medium at the heliospheric boundary.
\par
In this study, the CNN method effectively reduces the inversion error under short staring time by learning the complex mapping between the detector response $\boldsymbol{D}$ and the incident source $\boldsymbol{S}$. However, the application scenario of the currently trained CNN model still does not fully cover the data with extremely low flux, neither have we tested the performance with high-frequency temporal fluctuations. In the future, we need to further optimize the CNN model, enriching the training set to include cases with longer staring time or some other extreme cases and optimizing the learning parameters to improve the adaptability of the model to different application scenarios and support the long-term operation of HERMA.
\par
\par
\textbf{
NASA's Interstellar Mapping and Acceleration Probe (IMAP) carries three 
complementary ENA imagers, IMAP-Lo, IMAP-Hi, and IMAP-Ultra, covering broad 
and partially overlapping energy ranges. IMAP-Ultra measures ENAs over 
approximately 3--300 keV, which substantially overlaps the HERMA energy range 
of 10--500 keV, with IMAP-Ultra extending to lower energies and HERMA extending 
to higher energies. IMAP-Ultra provides access to most of the sky through its 
wide-field viewing geometry and spacecraft rotation, while the several-month 
cadence of its global map products mainly reflects the accumulation time 
required to obtain adequate counting statistics.}
\textbf{
Under the ENA flux, exposure, detector-response, and background assumptions 
adopted in the present simulations, HERMA demonstrates the potential to 
reconstruct near-global $\sim10$ keV ENA distributions on a cadence of several 
days. This projected high-cadence capability results from the combination of 
the spatial modulation configuration and CNN-based inversion, rather than from 
broader instantaneous sky accessibility than IMAP-Ultra. The actual in-flight 
mapping cadence will depend on the instrument efficiency, background level, 
spacecraft orbit, pointing strategy, solar-avoidance constraints, and the 
measured ENA flux. HERMA could therefore provide valuable high-cadence 
observations complementary to IMAP-Ultra in their overlapping energy range. 
Future contemporaneous IMAP observations may also provide an independent 
reference for cross-comparing the large-scale ENA morphology reconstructed by 
HERMA, while quantitative validation of the HERMA inversion will require 
dedicated calibration and end-to-end response characterization.}

\section*{Acknowledgments}


\subsection*{Author Contributions} 
W.-N. Wang and R. Lin contributed equally to this work as co-first authors. J.-S. He and H. Zou proposed the original idea and research framework of this study. W.-N. Wang and R. Lin performed the Geant4 simulations, and W.-N. Wang developed the CNN inversion model. All authors participated in result interpretation, manuscript discussion, and critical revision. W.-N. Wang, R. Lin, and J. He drafted the initial manuscript, and all authors approved the final version for publication.





\subsection*{Funding}
The work at Peking University was supported by CNSA, NSFC, and NKRDPC under Grant Nos. of D050106, D010202, D010301, 42530105, 42241118,  42150105, 2021YFA0718600, and 2022YFF0503800.




\subsection*{Conflicts of Interest}
There are no competing interests.



\subsection*{Data Availability}


The simulation datasets generated and analyzed in this study are available from the authors upon reasonable request.
\printbibliography

@article{kraft1988software,
  title={A software package for sequential quadratic programming},
  author={Kraft, Dieter},
  journal={Forschungsbericht- Deutsche Forschungs- und Versuchsanstalt fur Luft- und Raumfahrt},
  year={1988}
}

@article{kornbleuth2018globally,
  title={Globally distributed energetic neutral atom maps for the “Croissant” heliosphere},
  author={Kornbleuth, Marc and Opher, Merav and Michael, Adam T and Drake, James F},
  journal={The Astrophysical Journal},
  volume={865},
  number={2},
  pages={84},
  year={2018},
  publisher={IOP Publishing}
}

@article{baliukin2020heliospheric,
  title={Heliospheric energetic neutral atoms: non-stationary modelling and comparison with IBEX-Hi data},
  author={Baliukin, II and Izmodenov, VV and Alexashov, DB},
  journal={Monthly Notices of the Royal Astronomical Society},
  volume={499},
  number={1},
  pages={441--454},
  year={2020},
  publisher={Oxford University Press}
}

@article{baliukin2022energetic,
  title={Energetic pickup proton population downstream of the termination shock as revealed by IBEX-Hi data},
  author={Baliukin, II and Izmodenov, VV and Alexashov, DB},
  journal={Monthly Notices of the Royal Astronomical Society},
  volume={509},
  number={4},
  pages={5437--5453},
  year={2022},
  publisher={Oxford University Press}
}

@article{dayeh2022evolution,
  title={Evolution of the Heliotail Lobes over a Solar Cycle as Measured by IBEX},
  author={Dayeh, Maher A and Zirnstein, Eric J and Fuselier, Stephen A and Funsten, Herbert O and McComas, David J},
  journal={The Astrophysical Journal Supplement Series},
  volume={261},
  number={2},
  pages={27},
  year={2022},
  publisher={IOP Publishing}
}

@article{fuselier2009width,
  title={Width and variation of the ENA flux ribbon observed by the Interstellar Boundary Explorer},
  author={Fuselier, SA and Allegrini, F and Funsten, HO and Ghielmetti, AG and Heirtzler, D and Kucharek, H and Lennartsson, OW and McComas, DJ and M{\"o}bius, E and Moore, TE and others},
  journal={science},
  volume={326},
  number={5955},
  pages={962--964},
  year={2009},
  publisher={American Association for the Advancement of Science}
}

@article{funsten2009structures,
  title={Structures and spectral variations of the outer heliosphere in IBEX energetic neutral atom maps},
  author={Funsten, HO and Allegrini, F and Crew, GB and DeMajistre, R and Frisch, PC and Fuselier, SA and Gruntman, M and Janzen, P and McComas, DJ and M{\"o}bius, E and others},
  journal={Science},
  volume={326},
  number={5955},
  pages={964--966},
  year={2009},
  publisher={American Association for the Advancement of Science}
}

@article{mccomas2009global,
  title={Global observations of the interstellar interaction from the Interstellar Boundary Explorer (IBEX)},
  author={McComas, DJ and Allegrini, F and Bochsler, P and Bzowski, M and Christian, ER and Crew, GB and DeMajistre, R and Fahr, H and Fichtner, H and Frisch, PC and others},
  journal={Science},
  volume={326},
  number={5955},
  pages={959--962},
  year={2009},
  publisher={American Association for the Advancement of Science}
}

@article{dialynas2017bubble,
  title={The bubble-like shape of the heliosphere observed by Voyager and Cassini},
  author={Dialynas, K and Krimigis, SM and Mitchell, DG and Decker, RB and Roelof, EC},
  journal={Nature Astronomy},
  volume={1},
  number={5},
  pages={0115},
  year={2017},
  publisher={Nature Publishing Group UK London}
}

@article{krimigis2009imaging,
  title={Imaging the interaction of the heliosphere with the interstellar medium from Saturn with Cassini},
  author={Krimigis, SM and Mitchell, DG and Roelof, EC and Hsieh, KC and McComas, DJ},
  journal={Science},
  volume={326},
  number={5955},
  pages={971--973},
  year={2009},
  publisher={American Association for the Advancement of Science}
}

@inproceedings{krimigis2010ena,
  title={ENA (E> 5 keV) Images from Cassini and Voyager “ground truth”: Suprathermal Pressure in the Heliosheath},
  author={Krimigis, SM and Mitchell, DG and Roelof, EC and Decker, RB},
  booktitle={AIP Conference Proceedings},
  volume={1302},
  number={1},
  pages={79--85},
  year={2010},
  organization={American Institute of Physics}
}

@article{izmodenov2009local,
  title={Local interstellar parameters as they are inferred from analysis of observations inside the heliosphere},
  author={Izmodenov, Vladislav V},
  journal={Space science reviews},
  volume={143},
  number={1},
  pages={139--150},
  year={2009},
  publisher={Springer}
}

@article{zank2013heliospheric,
  title={Heliospheric structure: the bow wave and the hydrogen wall},
  author={Zank, GP and Heerikhuisen, J and Wood, Be E and Pogorelov, NV and Zirnstein, E and McComas, DJ},
  journal={The Astrophysical Journal},
  volume={763},
  number={1},
  pages={20},
  year={2013},
  publisher={The American Astronomical Society}
}

@article{mccomas2021interstellar,
  title={Interstellar pickup ion observations halfway to the termination shock},
  author={McComas, DJ and Swaczyna, P and Szalay, JR and Zirnstein, EJ and Rankin, JS and Elliott, HA and Singer, K and Spencer, J and Stern, SA and Weaver, H},
  journal={The Astrophysical Journal Supplement Series},
  volume={254},
  number={1},
  pages={19},
  year={2021},
  publisher={IOP Publishing}
}

@article{mccomas2024fourteen,
  title={Fourteen years of energetic neutral atom observations from IBEX},
  author={McComas, DJ and Alimaganbetov, M and Beesley, LJ and Bzowski, M and Funsten, HO and Janzen, PH and Kubiak, MA and Rankin, JS and Reisenfeld, DB and Schwadron, Nathan A and others},
  journal={The Astrophysical Journal Supplement Series},
  volume={270},
  number={2},
  pages={17},
  year={2024},
  publisher={IOP Publishing}
}

@article{mccomas2018interstellar,
  title={Interstellar mapping and acceleration probe (IMAP): A new NASA mission},
  author={McComas, DJ and Christian, Eric R and Schwadron, Nathan A and Fox, N and Westlake, J and Allegrini, F and Baker, DN and Biesecker, D and Bzowski, M and Clark, G and others},
  journal={Space science reviews},
  volume={214},
  number={8},
  pages={116},
  year={2018},
  publisher={Springer}
}

@article{wang2025spectral,
  title={Spectral Evolution of Pickup Ions across Interplanetary Shocks in the Outer Heliosphere: Observational Constraints and Transport Simulations},
  author={Wang, Weining and He, Jiansen and Duan, Die and Yang, Zhongwei and Wang, Linghua and Li, Hui},
  journal={The Astrophysical Journal Letters},
  volume={988},
  number={2},
  pages={L66},
  year={2025},
  publisher={The American Astronomical Society}
}

@article{shrestha2024suprathermal,
  title={Suprathermal H+ pickup ion tails in the outer heliosphere},
  author={Shrestha, Bishwas L and Zirnstein, Eric J and McComas, David J and Brandt, Pontus and Stern, Alan and Elliott, Heather A and Poppe, Andrew R and Singer, Kelsi N and Verbiscer, Anne},
  journal={The Astrophysical Journal},
  volume={960},
  number={1},
  pages={35},
  year={2024},
  publisher={The American Astronomical Society}
}

@article{mccomas2017interstellar,
  title={Interstellar pickup ion observations to 38 au},
  author={McComas, DJ and Zirnstein, EJ and Bzowski, M and Elliott, HA and Randol, B and Schwadron, NA and Sok{\'o}{\l}, JM and Szalay, JR and Olkin, C and Spencer, J and others},
  journal={The Astrophysical Journal Supplement Series},
  volume={233},
  number={1},
  pages={8},
  year={2017},
  publisher={The American Astronomical Society}
}

@article{zirnstein2018situ,
  title={In situ observations of preferential pickup ion heating at an interplanetary shock},
  author={Zirnstein, EJ and McComas, David J and Kumar, Rahul and Elliott, Heather Alison and Szalay, Jamey R and Olkin, CB and Spencer, J and Stern, S Alan and Young, Leslie Ann},
  journal={Physical review letters},
  volume={121},
  number={7},
  pages={075102},
  year={2018},
  publisher={APS}
}

@article{kornbleuth2023probing,
  title={Probing the length of the heliospheric tail with energetic neutral atoms (ENAs) from 0.52 to 80 keV},
  author={Kornbleuth, M and Opher, M and Dialynas, K and Zank, GP and Wang, BB and Baliukin, I and Gkioulidou, M and Giacalone, J and Izmodenov, V and Sok{\'o}{\l}, JM and others},
  journal={The Astrophysical Journal Letters},
  volume={945},
  number={1},
  pages={L15},
  year={2023},
  publisher={The American Astronomical Society}
}

@article{zhang2024shape,
  title={The Shape of the Heliosphere Derived from the IBEX Ribbon},
  author={Zhang, Shuai and Liu, Kaijun and Shi, Quanqi and Tian, Anmin and Yao, Fei},
  journal={The Astrophysical Journal Letters},
  volume={977},
  number={2},
  pages={L39},
  year={2024},
  publisher={The American Astronomical Society}
}

@article{yu2025design,
  title={Design and calibration of the grid-based energetic neutral atom imager (GENA) for the Queqiao-2 mission},
  author={Yu, Q. and Sun, T. and Zhang, X. and others},
  journal={Science China Technological Sciences},
  volume={68},
  number={11},
  pages={2100603},
  year={2025},
  publisher={Science China Press}
}

@article{wang2025image,
  title={Image simulation of grid-based energetic neutral atom imager onboard the Queqiao-2 mission},
  author={Wang, D. and Lei, L. and Yu, X. and others},
  journal={Science China Technological Sciences},
  volume={68},
  number={11},
  pages={1--16},
  year={2025},
  publisher={Science China Press}
}

@article{burlaga2019magnetic,
  title={Magnetic field and particle measurements made by Voyager 2 at and near the heliopause},
  author={Burlaga, LF and Ness, NF and Berdichevsky, DB and Park, J and Jian, LK and Szabo, A and Stone, EC and Richardson, JD},
  journal={Nature Astronomy},
  volume={3},
  number={11},
  pages={1007--1012},
  year={2019},
  publisher={Nature Publishing Group UK London}
}

@article{zirnstein2017structure,
  title={Structure of the heliotail from interstellar boundary explorer observations: implications for the 11-year solar cycle and pickup ions in the heliosheath},
  author={Zirnstein, EJ and Heerikhuisen, J and Zank, GP and Pogorelov, NV and Funsten, Herbert O and McComas, DJ and Reisenfeld, Daniel Brett and Schwadron, NA},
  journal={The Astrophysical Journal},
  volume={836},
  number={2},
  pages={238},
  year={2017},
  publisher={The American Astronomical Society}
}

@misc{siegl2009standard,
  title={Standard radiation environment monitor: simulation and inner belt flux anisotropy investigation},
  author={Siegl, Martin},
  year={2009}
}

@article{hurford2002rhessi,
  title={The RHESSI imaging concept},
  author={Hurford, GJ c and Schmahl, EJ and Schwartz, RA and Conway, AJ and Aschwanden, MJ and Csillaghy, A and Dennis, BR and Johns-Krull, C and Krucker, S and Lin, RP and others},
  journal={Solar Physics},
  volume={210},
  number={1},
  pages={61--86},
  year={2002},
  publisher={Springer}
}

@article{massa2023stix,
  title={The STIX imaging concept},
  author={Massa, Paolo and Hurford, Gordon J and Volpara, Anna and Kuhar, Matej and Battaglia, Andrea F and Xiao, Hualin and Casadei, Diego and Perracchione, Emma and Garbarino, Sara and Guastavino, Sabrina and others},
  journal={Solar Physics},
  volume={298},
  number={10},
  pages={114},
  year={2023},
  publisher={Springer}
}

@article{chen2021design,
  title={Design and verification of the HXI collimator on the ASO-S mission},
  author={Chen, Deng-Yi and Hu, Yi-Ming and Ma, Tao and Su, Yang and Yang, Jian-Feng and Wang, Jian-Ping and Xu, Guang-Zhou and Jiang, Xian-Kai and Guo, Jian-Hua and Zhang, Yong-Qiang and others},
  journal={Research in Astronomy and Astrophysics},
  volume={21},
  number={6},
  pages={136},
  year={2021},
  publisher={National Astronomical Observatories, CAS and IOP Publishing Ltd.}
}

@incollection{funsten2009ibexhi,
  title={The Interstellar Boundary Explorer High Energy (IBEX-Hi) Neutral Atom Imager},
  author={Funsten, Herbert O. and Allegrini, Frederic and Bochsler, Peter and Chornay, David and DeMajistre, Robert and Gloeckler, George and Janzen, Paul H. and McComas, David J. and Möbius, Eberhard and Rodriguez, Manuel and others},
  booktitle={Interstellar Boundary Explorer (IBEX)},
  pages={75--103},
  year={2009},
  publisher={Springer},
  address={New York, NY}
}

@article{fuselier2009ibexlo,
  title={The IBEX-Lo Sensor},
  author={Fuselier, Stephen A. and Bochsler, Peter and Chornay, David and Clark, Gregory and Dunn, Geoffrey and Fok, Mei-Ching and Funsten, Herbert O. and Ghielmetti, Arthur G. and Goldstein, Jerry and Janzen, Paul and others},
  journal={Space Science Reviews},
  volume={146},
  pages={117--147},
  year={2009},
  publisher={Springer}
}

@article{schwadron2018heliosphere,
  title={The Heliosphere Is Not Round},
  author={Schwadron, Nathan A. and Bzowski, Maciej},
  journal={The Astrophysical Journal},
  volume={862},
  number={1},
  pages={11},
  year={2018},
  publisher={IOP Publishing},
  doi={10.3847/1538-4357/aacbcf}
}

@article{Orsini2010SERENA,
  author  = {Orsini, S. and Livi, S. and Torkar, K. and others},
  title   = {{SERENA}: A suite of four instruments ({ELENA}, {STROFIO}, {PICAM} and {MIPA}) on board {BepiColombo-MPO} for particle detection in the Hermean environment},
  journal = {Planetary and Space Science},
  year    = {2010},
  volume  = {58},
  number  = {1--2},
  pages   = {166--181}
}

@article{chen2026ground,
  title={Ground calibration of the helios ENA Rapid MApper (HERMA) for heliospheric boundary exploration},
  author={Chen, Jia-Li and Zou, Hong and He, Jian-Sen and Ye, Yu-Guang and Sun, Jia-Chen and Wang, Wei-Ning and Shi, Wei-Hong and Wang, Yong-Fu and Lin, Rong and Chen, Tian-Hang and others},
  journal={Journal of Geophysical Research: Space Physics},
  volume={131},
  number={9},
  pages={e2026JA035573},
  year={2026},
  publisher={Wiley Online Library}
}

\end{document}